\documentclass[reprint, superscriptaddress, secnumarabic, amssymb, nobibnotes, aps, prl]{revtex4-1}

\usepackage{graphicx}
\usepackage{epstopdf}
\usepackage[T1]{fontenc}
\usepackage[latin9]{inputenc}
\usepackage{amsbsy}
\usepackage{gensymb}
\usepackage[T1]{fontenc}
\usepackage[latin9]{inputenc}
\usepackage{amsmath}
\usepackage{amssymb}
\usepackage{bbm}
\usepackage{physics}
\usepackage{braket}
\usepackage{xcolor}
\allowdisplaybreaks
\usepackage{graphicx}
\usepackage[colorlinks=true]{hyperref}  
\hypersetup{
    bookmarks=true,         % show bookmarks bar?
    unicode=false,          % non-Latin characters 
    pdftoolbar=true,        % show Acrobat
    pdfmenubar=true,        % show Acrobat 
    pdffitwindow=false,     % window fit to page when opened
    pdfstartview={FitH},    % fits the width of the page to the window
    pdftitle={Observation of time reversal symmetry breaking in noncentrosymmetric superconductor La7Rh3},    % title
    pdfauthor={Arushi, Singh, Hillier, Scheurer, Singh},     % author
    pdfsubject={},   % subject of the document
    pdfcreator={},   % creator of the document
    pdfproducer={}, % producer of the document
    pdfkeywords={} {} {}, % list of keywords
    pdfnewwindow=true,      % links in new window
    colorlinks=true,       % false: boxed links; true: colored links
    linkcolor=blue, %red,          % color of internal links (change box color with linkbordercolor)
    citecolor=blue,        % color of links to bibliography
    filecolor=magenta,      % color of file links
    urlcolor=blue           % color of external links
} 
\usepackage[normalem]{ulem}

\newcommand{\equref}[1]{Eq.~(\ref{#1})}

\newcommand{\figref}[1]{Fig.~\ref{#1}}
\newcommand{\refcite}[1]{Ref.~\onlinecite{#1}}

\newcommand{\appref}[1]{Appendix~\ref{#1}}

\renewcommand{\approx}{\simeq}

\renewcommand{\vec}[1]{\boldsymbol{#1}}

\begin{document}
\title{Time-reversal symmetry breaking and multigap superconductivity in the noncentrosymmetric superconductor La$_{7}$Ni$_{3}$}

\author{Arushi}
\affiliation{Department of Physics, Indian Institute of Science Education and Research Bhopal, Bhopal, 462066, India}
\author{D.~Singh}
\affiliation{ISIS Facility, STFC Rutherford Appleton Laboratory, Didcot OX11 0QX, United Kingdom}
\author{A.~D.~Hillier}
\affiliation{ISIS Facility, STFC Rutherford Appleton Laboratory, Didcot OX11 0QX, United Kingdom}
\author{M.~S.~Scheurer}
\affiliation{Institute for Theoretical Physics, University of Innsbruck, A-6020 Innsbruck, Austria}
\author{R.~P.~Singh}
\email[]{rpsingh@iiserb.ac.in}
\affiliation{Department of Physics, Indian Institute of Science Education and Research Bhopal, Bhopal, 462066, India}
\date{\today}
\begin{abstract}
The Th$_{7}$Fe$_{3}$ family of superconductors provides a rich playground for unconventional superconductivity. La$_7$Ni$_3$ is the latest member of this family, which we here investigate by means of thermodynamic and muon spin rotation and relaxation measurements. 
Our specific heat data provides evidence for two distinct and approximately isotropic superconducting gaps. The larger gap has a value slightly higher than that of weak-coupling BCS theory, indicating the presence of significant correlations. These observations are confirmed by transverse-field muon-rotation measurements.
Furthermore, zero-field measurements reveal small internal fields in the superconducting state, which occur close to the onset of superconductivity and indicate that the superconducting order parameter breaks time-reversal symmetry. 
We discuss two possible microscopic scenarios---an unconventional $E_{2}(1,i)$ state and an $s+i\,s$ superconductor, which is reached by two consecutive transitions---and illustrate which interactions will favor these phases. 
Our results establish La$_{7}$Ni$_{3}$ as the first member of the Th$_{7}$Fe$_{3}$ family displaying both time-reversal-symmetry-breaking and multigap superconductivity.
\end{abstract}
\maketitle

% Symmetry breaking in general and inversion asymmetric systems with spin-orbit coupling:
While the superconducting transition of the original Bardeen-Cooper-Schrieffer (BCS) theory is not a symmetry-breaking phase transition, but rather a condensed-matter realization of a Higgs mechanism, modern research in superconductivity is crucially concerned with aspects of symmetry \cite{Sigrist,Mix4}: first, the superconducting order parameter might transform non-trivially under a subgroup of the symmetries of the normal state. In that case, the superconducting transition is also a symmetry-breaking phase transition. Second, a lot of research is concerned with the realization and study of superconductivity in systems with reduced symmetries, which is driven by the goal of designing superconductors with exotic and potentially useful properties. Among these, non-centrosymmetric crystal structures with significant spin-orbit coupling are of particular interest \cite{Mix4}; they exhibit inversion anti-symmetric Rashba-Dresselhaus spin-orbit-coupling terms \cite{Rashba,Dresselhaus}, which result in the splitting of the Fermi surfaces into two opposite spin configurations. 
This has a pronounced effect on the possible pairing states and leads to mixed-parity spin states (singlet and triplet) in the order parameter \cite{Mix1,Mix2,Mix3,Mix4}. As a result, it can give rise to some novel and uncommon superconducting properties, such as unusually high upper critical fields, larger than the Pauli limit \cite{Pauli1,Pauli2,Pauli3,Pauli4,Pauli5}, the presence of non-trivial line or point nodes \cite{line1,line2,line3,line4,line5}, multiple-gap behavior \cite{multi1,multi2}, an enhanced protection against impurity scattering \cite{DisorderSOCFu,OurDisorderSOC,BrydonScattering,PdTeScattering,Jonathan,Ando2012,Ando2014,Welp}, and the more recently proposed topological superconductivity \cite{topo0,topo1,topo2,topo3,topo4,topo5}.

% Time-reversal symmetry and why cool: 
One of the most fascinating and intensely-studied cases of broken symmetries in the superconducting state is time-reversal-symmetry (TRS) breaking. This is not only due to the fact that TRS-breaking superconductors can exhibit topological chiral edge modes \cite{topo0}, but also by the fundamental role of TRS for superconductivity itself \cite{Anderson}, which is associated with the formation of bound states of electrons at momenta related by time-reversal.
While a lot of weakly correlated non-centrosymmetric superconductors have been studied \cite{LPtPdSi3,Mo3P,Mo3Rh2N,LPG}, TRS breaking has only been observed in a handful of them \cite{Y5R6S18,LNC,SPA,R6Z,R6H,R6T,R24T5,L7I3_TRSB,Re}. 
In general, the key ingredients for TRS breaking at a superconducting transition are mostly unknown, although some general constraints have recently been derived theoretically for spin-orbit-coupled non-centrosymmetric systems \cite{DesignPrinciples,SheurerGenRelation}.

% Next, some specific materials:
Non-centrosymmetric superconductivity in compounds with $\alpha$-Mn structure (with or without Re concentration) has been examined a lot; however, indications of TRS breaking have only been observed for Re-based superconductors, where the magnitude of the internal field is proportional to the Re concentration but insensitive to other elements in the alloy \cite{R6Z,R6H,R6T,R24T5}. 
The relevance of Re is corroborated further by theoretical calculations \cite{R6Z_band} that suggest the dominance of the density of states of Re at the Fermi level. These observations are indicative of the vital role played by Re together with the lack of inversion symmetry for the formation of superconductivity.

\begin{figure*}[t]
\includegraphics[width=2.0\columnwidth,origin=b]{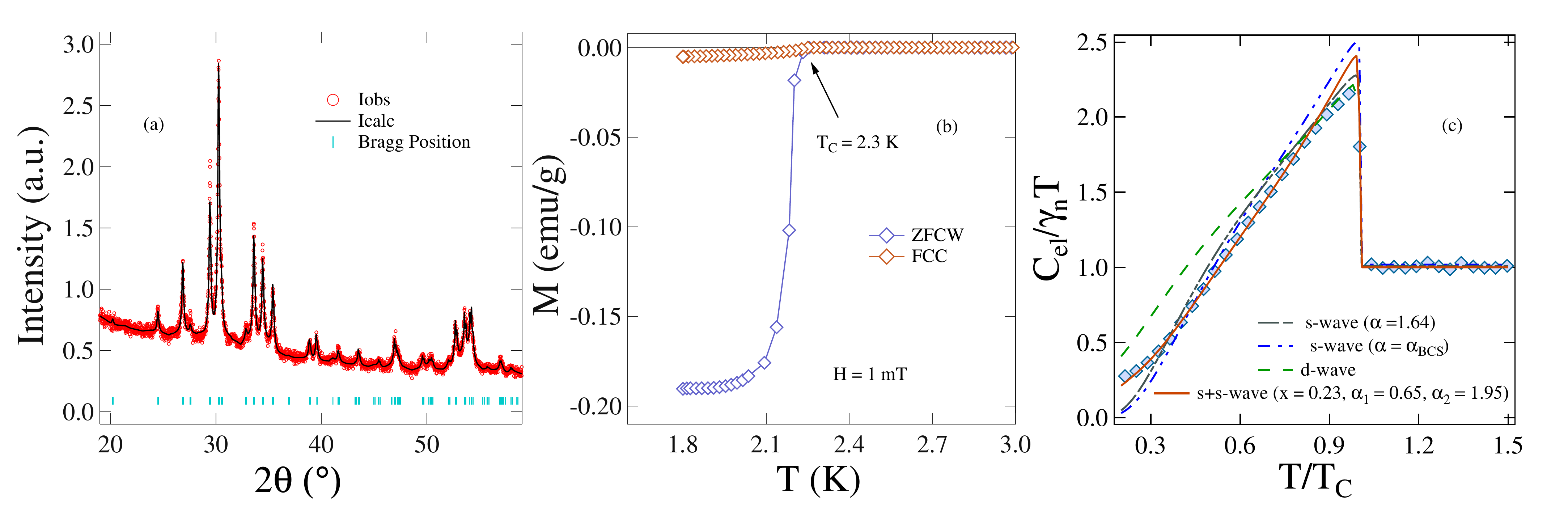}
\caption{\label{Fig1:xrd} (a) Powder XRD pattern of La$_{7}$Ni$_{3}$ is presented by red dots whereas the black line corresponds to the Rietveld refinement. (b) Magnetization as a function of temperature collected in the zero-field cooled warming (ZFCW) and field cooled cooling (FCC), confirming superconductivity with critical temperature $T_{c}$ = 2.3 K. (c) Temperature dependence of normalized electronic specific heat data, where the dashed and solid lines represent fits to the different models described in the main text.}
\end{figure*}

% Now specifically coming to our system:  
Similarly, the study of other non-centrosymmetric superconductors with a heavy element (other than Re) whose density of states at the Fermi level dominates can help us establish a relation between broken inversion symmetry, TRS of the superconductor, and the existence of a dominant density of states of a particular element in a compound. 
In this regard, the non-centrosymmetric superconducting family La$_{7}$X$_{3}$ (X = Ir, Rh, Ni), crystallizing in a Th$_{7}$Fe$_{3}$-type structure, which already shows examples of TRS breaking \cite{L7I3_TRSB,L7R3_TRSB}, provides potential candidates to investigate the above suggestions. 
In these cases also, the magnitude of the TRS-breaking signal remains unchanged after replacing a 5d [La$_{7}$Ir$_{3}$] \cite{L7I3_TRSB} by a 4d element [La$_{7}$Rh$_{3}$] \cite{L7R3_TRSB}, like in the Re$_{6}$X series compounds \cite{R6Z,R6H,R6T,R24T5}. 
This may be due to the dominant role played by Re and La at the respective Fermi surfaces \cite{R6Z_band,L7I3_band}.

In this paper, the properties of La$_{7}$Ni$_{3}$, the youngest member of the La$_{7}$X$_{3}$ series, are analyzed using muon spin relaxation and rotation ($\mu$SR) supplemented by thermodynamic measurements. $\mu$SR in zero-field mode is one of the best-suited techniques for studying TRS breaking as it is susceptible to tiny changes in the internal fields ($\approx10\mu$T). Our zero-field measurements reveal the appearance of spontaneous internal magnetic fields in the superconducting state. This provides substantial evidence for broken TRS in La$_{7}$Ni$_{3}$, with magnitude similar to La$_{7}$Ir$_{3}$ and La$_{7}$Rh$_{3}$. Generally, the spin-orbit coupling contribution from a 3d element, such as Ni, is considered negligible, which points towards the probable dominance of La in determining the superconducting ground state of La$_{7}$Ni$_{3}$. 
We further perform $\mu$SR measurements in the presence of a transverse magnetic field, which allows us to determine the temperature evolution of the magnetic penetration depth ($\lambda$) and, thus, infers additional information on the superconducting gap of La$_{7}$Ni$_{3}$.

\section{Experimental details}

The sample of La$_{7}$Ni$_{3}$ was prepared by arc melting stoichiometric amounts of La (99.9\%) and Ni (99.95\%) in a high purity argon gas atmosphere on a water-cooled copper hearth. The button was melted several times to ensure phase homogeneity with negligible weight loss (< 1\%). Magnetization and specific heat measurements were carried out using Quantum design MPMS 3 and physical property measurement system (PPMS).  

Muon-spin rotation and relaxation measurements were performed at the ISIS pulsed neutron and muon source, Rutherford Appleton Laboratory, the United Kingdom in two different configurations (transverse, zero/longitudinal field). A detailed account on the $\mu$SR technique can be found in \cite{Muon}. Three sets of orthogonal coils and an active compensation system are used to cancel the stray fields within  1 $\mu$T, which can be present at the sample position due to the Earth and neighboring instruments. La$_{7}$Ni$_{3}$ in powdered form was placed in a dilution refrigerator after mounting on a silver holder, with diluted GE varnish, which is operated in the temperature range 0.1 K $\leq$ T $\leq$ 3.0 K.

\section{Results}

The powder X-ray diffraction (XRD) collected at room temperature, shown in \figref{Fig1:xrd}(a) confirmed the crystal structure as hexagonal with space group $P6_{3}mc$ (No.~186). The lattice constants are: $a = b = 10.143 \pm 0.010$ \text{\AA} and $c = 6.468 \pm 0.006$ \text{\AA} which are in agreement with the previously published data \cite{L7N3_crys}. From the magnetization measurements shown in \figref{Fig1:xrd}(b) we find a superconducting transition temperature of $T_{c}$ = 2.3 $\pm$ 0.1~K, which is in agreement with previously published results \cite{L7N3_macro} and is found to be accompanied by a drop in resistivity (see \appref{ExtractParameter}). The superconducting parameters coherence length, penetration depth, and critical fields of La$_{7}$Ni$_{3}$ have been extracted from our magnetization/susceptibility measurements and are summarized in Table~\ref{TableWithParameters}.

\subsection{Specific heat}
We have also measured the specific heat $C$. To extract the electronic part of it, $C_{\text{el}}$, we subtracted the phononic contribution from the total heat capacity, $C_{\text{el}}  = C - C_{\text{ph}} = C - \beta_{3} T^{3} + \beta_{5}T^{5}$ where $\beta_{3}$ and $\beta_{5}$, characterizing the phononic contribution, are obtained by fitting them to the data. $C_{\text{el}}$, normalized with respect to $\gamma_n T$ (with Sommerfeld coefficient $\gamma_n$), is shown in \figref{Fig1:xrd}(c) as a function of reduced temperature, $T/T_{c}$. The data of $C_{\text{el}}$ were fitted to the following models: 
(i) a single $s$-wave gap, which is taken to be isotropic and where $\alpha=\Delta (T=0)/k_{B}T_{c}$ is set to the BCS value, $\alpha=\alpha_{\text{BCS}} \approx 1.764$; (ii) an isotropic $s$-wave gap with $\alpha$ as a variable parameter, also known as the $\alpha$-model \cite{AlphaModel}; (iii) a line-nodal $d$-wave gap; (iv) a phenomenological two-gap $\alpha$-model \cite{2gapModel} with three fitting parameters---the values of $\alpha_j = \Delta_j (0)/k_{B}T_{c}$ of the two gaps, and $x$ (with $0\leq x\leq 0.5$ without loss of generality), which parameterizes the ratio of the partial Sommerfeld coefficients (or densities of states), $\gamma_1=x\gamma_n$ and $\gamma_2=(1-x)\gamma_n$, of the parts of the Fermi surface with gap values $\Delta_1$ and $\Delta_2$, respectively.

The resulting fits for the different models are shown in \figref{Fig1:xrd}(c). We can clearly see that the medium- and low-temperature behavior is captured only by the two-gap model. Fitting yields $x = 0.23$, $\alpha_{1}$ = 0.65 and $\alpha_{2}$ = 1.95.
%The results of the different analysis are represented by the solid and dashed lines in \figref{Fig1:xrd}(c). As can be seen in \figref{Fig1:xrd}(c), the single gap with $\alpha$ either constant or variable does not describe the data sufficiently. The excellent fit to the data is obtained using the two-gap model which yield the parameters as: $x = 0.23$, $\alpha_{1}$ = 0.65 and $\alpha_{2}$ = 1.95. 
We, thus, find a quite large difference in gap magnitude with ratio $\Delta_{2}(0)/\Delta_{1}(0) \approx 3$. As expected, the relative weight of the smaller gap is suppressed, $x : (1-x) \approx 23\,\%:77\,\%$.
%The contribution from the small gap, $\alpha_{1}$ is very low together with its weight percentage suggests that the superconductivity in La$_{7}$Ni$_{3}$ is mainly carried out by the band with a large energy gap. 
The value of $\alpha_{2}$, which represents the larger gap, is a bit higher than the BCS value, corresponding to moderately coupled superconductivity in La$_{7}$Ni$_{3}$. 
These findings provide evidence that La$_{7}$Ni$_{3}$ might be a two-gap superconductor where both gaps open up at a common $T_{c}$ and have a similar temperature dependence. Further experiments and theoretical band structure calculations are required to shed more light on the two-gap feature in La$_{7}$Ni$_{3}$.

\begin{figure} 
\includegraphics[width=1.0\columnwidth]{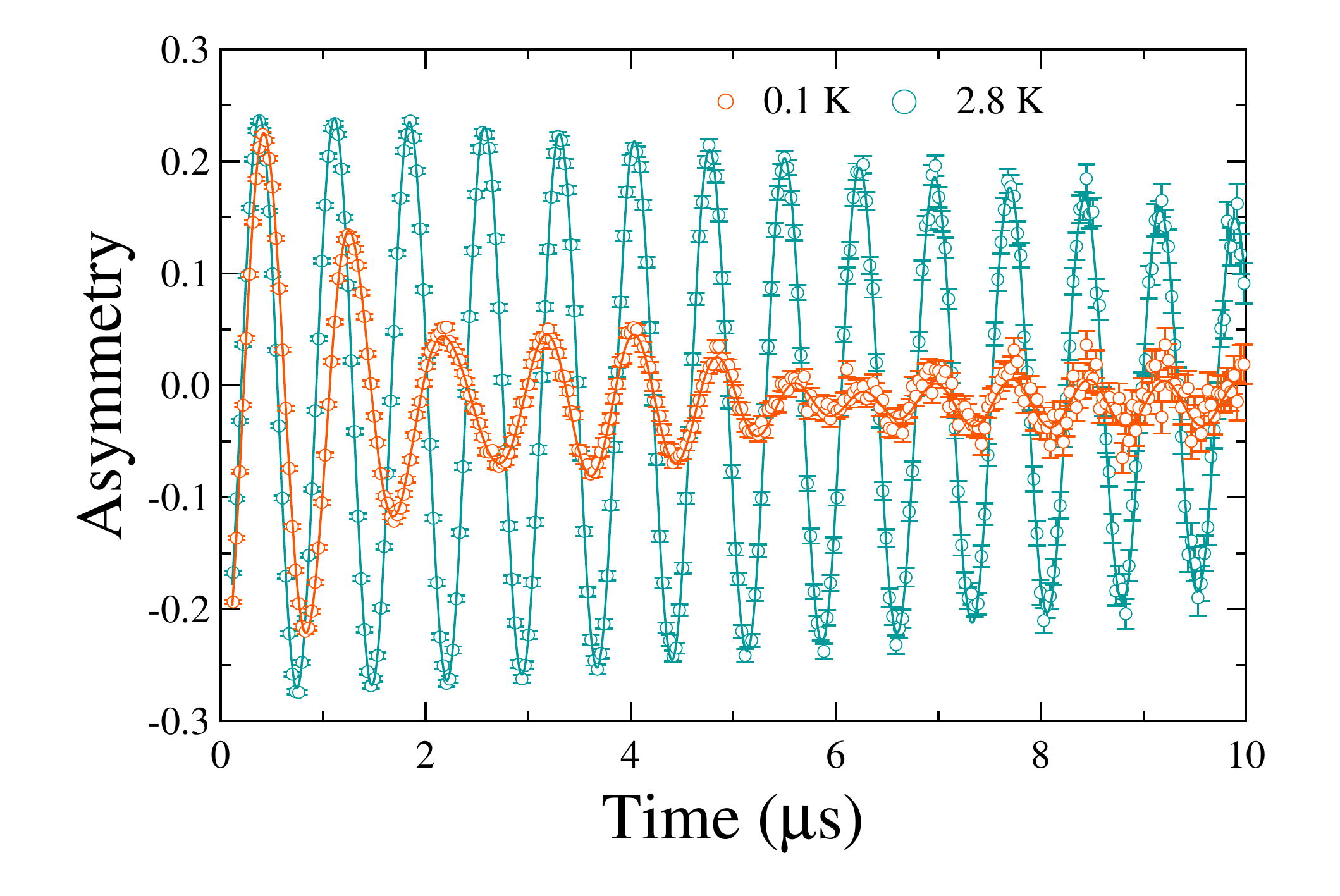} %{r}{0.5\textwidth}
\caption{\label{Fig2:TF_Asymm} The time-domain TF spectra taken above and below $T_{c}$ in an applied magnetic field of 10 mT. The solid lines show the fitting using \equref{eqn1:Tranf}.}
\end{figure}

\begin{figure*}
\includegraphics[width=2.0\columnwidth,origin=b]{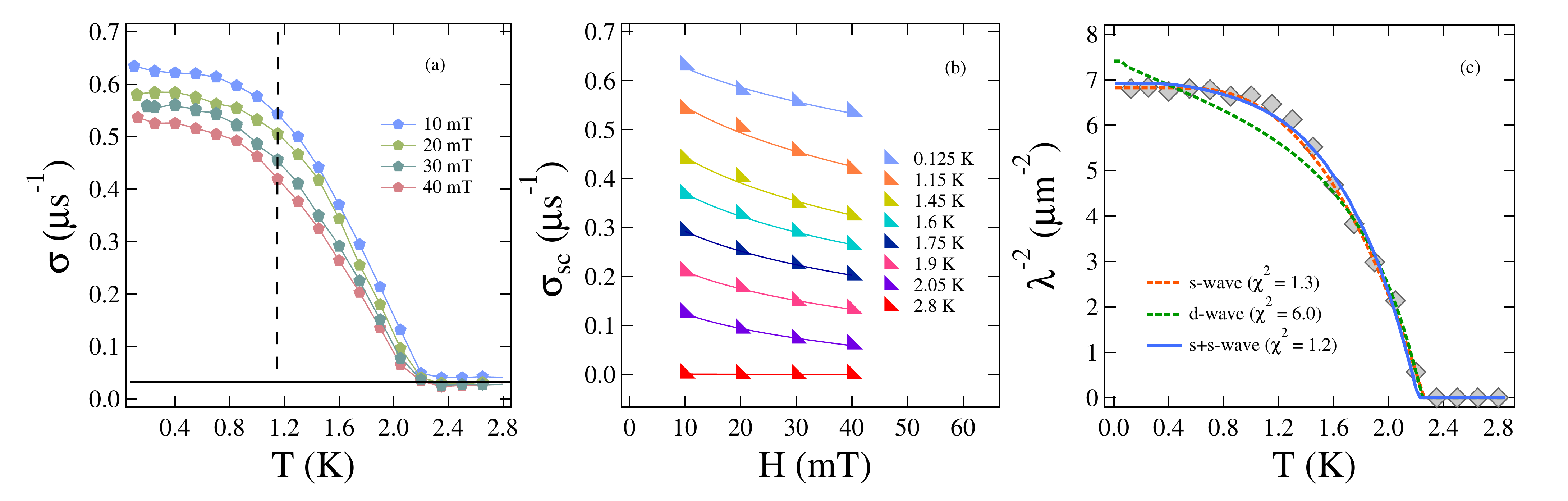}
\caption{\label{Fig3:TF} (a) Temperature dependence of the muon spin depolarization rate at different applied magnetic fields starting from 10 mT to 40 mT in TF-$\mu$SR measurements. (b) $\sigma_{sc}(H)$ at various temperatures ($T = 0.125\,\textrm{K}$ to $T = 2.80\,\textrm{K}$). The data were fitted using \equref{eqn3:sigmaH} to extract the temperature dependence of the inverse square of magnetic penetration depth. (c) The calculated $\lambda^{-2}$(T) where the solid line represents the best fit using \equref{eqn4:dirty_2gap}.}
\end{figure*}

\subsection{Transverse-field $\mu$SR and penetration depth}
To obtain complementary information on the nature of the superconducting gap, transverse-field (TF) $\mu$SR measurements were performed on La$_{7}$Ni$_{3}$. Magnetic fields in the range 10 mT to 40 mT, well above the lower critical field and, perpendicular to the initial muon spin direction were applied before cooling it through $T_{c}$ to 0.1 K. This generates a well-ordered flux line lattice in the mixed superconducting state of a type II superconductor, resulting in a distinctive field distribution throughout the sample. The muon asymmetry spectra for temperatures above and below $T_{c}$ are shown in \figref{Fig2:TF_Asymm}. For $T > T_{c}$, spectra show a weak depolarization rate because of the nuclear dipolar fields and also confirm the homogeneous field distribution throughout the sample. In contrast, below $T_{c}$, the pronounced Gaussian damping attributes to the inhomogeneous field distribution in the flux line lattice. The time-domain spectra are best described by the sum of sinusoidally oscillating functions, each damped with a Gaussian relaxation component \cite{TFfunc1,TFfunc2},
\begin{equation}
\begin{split}
A (t) = \sum_{i=1}^N A_{i}\exp\left(-\frac{1}{2}\sigma_i^2t^2\right)\cos(\gamma_\mu B_it+\phi)\\ + A_{bg}\cos(\gamma_\mu B_{bg}t+\phi),
\label{eqn1:Tranf}
\end{split}
\end{equation}
where $\phi$ is the initial phase, $A_{i}$ is the initial asymmetry, $B_{i}$ is the mean-field of the $i$th component of the Gaussian distribution, $\sigma_{i}$ is the relaxation rate, and $\gamma_{\mu}/2\pi$ = 135.5 MHz/T is the muon gyromagnetic ratio. $A_{bg}$ and $B_{bg}$ are the temperature-independent background contribution for asymmetry and field, respectively, which originates from the muon stopping at the sample holder. Two Gaussian components ($N=2$) are found sufficient to fit the time spectra for our sample. The second-moment method \cite{IImoment} is applied to calculate the temperature dependence of the total depolarization rate, $\sigma$, for different applied magnetic fields. The first and second moments of the local magnetic field distribution, $p(B)$, for $N = 2$ are given by \cite{IImoment}:
\begin{equation}
\expval{B} = \sum_{i=1}^2\frac{A_{i}B_{i}}{A_{1}+A_{2}},
\label{eqn2:Tranf}
\end{equation}
\begin{equation*}
\expval{\Delta B^2} = \frac{\sigma^2}{\gamma_{\mu}^2} = \sum_{i=1}^2\frac{A_{i}[(\sigma_{i}/\gamma_{\mu})^2+(B_{i}-\expval{B})^2]}{A_{1}+A_{2}},
\label{eqn2:Tranf}
\end{equation*}
where $\braket{\cdot}$ denotes average over magnetic fields.
Figure~\ref{Fig3:TF}(a) displays the resultant $\sigma(T)$. It has an almost constant value above $T_{c}$, which is due to the contribution of the nuclear dipolar field, $\sigma(T\gtrsim T_c) \approx\sigma_{\text{ndip}}$. The superconducting contribution, $\sigma_{\text{sc}}$, was extracted by subtracting $\sigma_{\text{ndip}}$ = 0.0623 $\mu$s$^{-1}$ from $\sigma$ using the following expression: $\sigma_{\text{sc}}$ = $(\sigma^{2} - \sigma_{\mathrm{\text{ndip}}}^{2})^{1/2}$. Figure \ref{Fig3:TF}(b) represents $\sigma_{\text{sc}}$, evaluated using isothermal cuts (dotted line) of the $\sigma(T)$ data sets in Fig.~\ref{Fig3:TF}(a). The magnetic-field dependence of the penetration depth, $\lambda_L$, for Ginzburg-Landau parameters $\kappa_{\text{GL}}$  $\ge$ 5 (we estimate $\kappa_{\text{GL}} \approx 26$, see \appref{ExtractParameter}) is given by \cite{sigmaSC}
\begin{equation}
\sigma_{\mathrm{sc}}(\mu s^{-1}) = 4.854 \times 10^{4}(1-b)[1+1.21(1-\sqrt{b})^{3}]\lambda^{-2} , 
\label{eqn3:sigmaH}
\end{equation}
for small reduced magnetic fields $b = \expval{B}/B_{c2}(T)$. \equref{eqn3:sigmaH} was fitted to each $\sigma_{\text{sc}}$ curve in order to extract the temperature dependence of $\lambda$. Figure \ref{Fig3:TF}(c) represents the behavior of the corresponding $\lambda^{-2}$(T). To gain information about the form of the superconducting gap from $\lambda^{-2}(T)$, we use that the latter can be expressed in the semi-classical approximation as \cite{SemiClassical,Multigap1,Multigap2}
\begin{equation}
\frac{\lambda^{-2}(T, \Delta_{0})}{\lambda^{-2}(0,\Delta_{0})} = 1+\frac{1}{\pi}\int_{0}^{2\pi}\int_{\Delta(T,\phi)}^{\infty}\frac{\partial f}{\partial E}\frac{EdEd\phi}{\sqrt{E^{2}-\Delta(T,\phi)^{2}}}, \\
\label{eqn7:BCS1}
\end{equation}
where we have, for concreteness, focused on an isotropic 3D system with a gap $\Delta(T,\phi)$ that only depends on the azimuthal angle $\phi$ and $\textit{f}$ = [exp($\textit{E}$/$k_{B}T$)+1]$^{-1}$ is the Fermi-Dirac function. We use $\Delta(T,\phi) = \Delta_{0}\delta(T/T_{c})g(\phi)$, where $\delta(T/T_{c}) = \tanh[1.82(1.018((\mathit{T_{c}/T})-1))^{0.51}$] describes (approximately) the temperature dependence of the overall gap magnitude, and $g(\phi)$ captures its angular dependence. We have $g(\phi)=1$ for an isotropic gap function; note that such a gap function is not only consistent with $s$-wave pairing, but also with the (single-gap limit of the) $E_2(1,i)$-wave and $s+i\,s$ pairing states discussed below. For a $d$-wave state, it holds $g(\phi)=|\cos(2\phi)|$, leading to nodal points for $\phi=\pm\pi/4,\pm 3\pi/4$.
To capture a possible two-gap scenario, we use a weighted sum \cite{2gap}
\begin{equation}
\frac{\lambda^{-2}(T)}{\lambda^{-2}(0)} = x\frac{\lambda^{-2}(T, \Delta_{0,1})}{\lambda^{-2}(0, \Delta_{0,1})}+(1-x)\frac{\lambda^{-2}(T, \Delta_{0,2})}{\lambda^{-2}(0, \Delta_{0,2})},
\label{eqn4:dirty_2gap}
\end{equation}
where $x \leq 0.5$ without loss of generality. %and we restrict ourselves to two isotropic gaps without any $\phi$ dependence in \equref{eqn7:BCS1}. 
Exactly as in the multi-gap $\alpha$-model for the specific heat above, the factor $x$ in \equref{eqn4:dirty_2gap} expresses the fact that the density of states (or Sommerfeld coefficients) of the parts of the Fermi surface with gap $\Delta_{0,1}$ and $\Delta_{0,2}$ are in general different ($x\neq 0.5$).

\begin{figure*}
\includegraphics[width=2.0\columnwidth,origin=b]{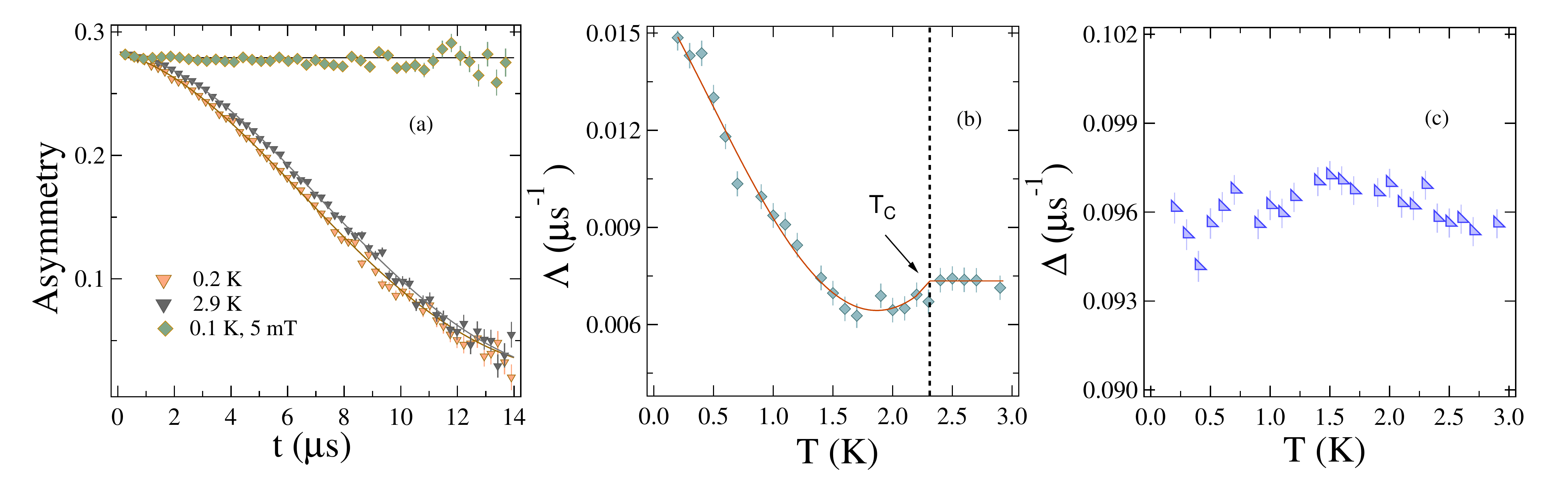}
\caption{\label{Fig4:ZF} (a) The asymmetry spectra collected at 0.2 K (orange) and 2.9 K (brown) in ZF-$\mu$SR, together with the LF-$\mu$SR data taken at 0.1 K in an applied field of 5mT (green). The solid lines are fits of \equref{eqn6:tay}. (b) $\Lambda(T)$ exhibits a significant increase below $T_c = 2.3 \,\textrm{K}$. (c) Temperature dependence of the nuclear relaxation rate $\Delta$, which shows no dependence on temperature.}  
\end{figure*}

The data shown in \figref{Fig3:TF}(c) was fitted with various models, consisting of a single isotropic gap (\equref{eqn7:BCS1} with $g(\phi)=1$), a single $d$-wave gap ($g(\phi)=|\cos(2\phi)|$ in \equref{eqn7:BCS1}), and a model with two isotropic gaps (\equref{eqn4:dirty_2gap} with $g(\phi)=1$). As can be seen, a nodal $d$-wave scenario is not consistent with the data, in accordance with our discussion of the specific heat. While both single- and two-gap $s$-wave models yield good agreement, the fit is slightly better for the latter ($\chi^2 = 1.2$). More importantly, the best fit is obtained for a quite significant imbalance of the gap magnitudes, $\Delta_{0,2}(0)$/$\Delta_{0,1}(0) \approx 2.7$, where the larger gap is hosted by the Fermi pocket(s) with the larger density of states---again, in line with the specific heat.
The absolute values, however, of the gaps, $\Delta_{0,1}(0)$ = 0.18 meV [$\Delta_{0,1}/k_{B}T_{c}$ = 0.94], and $\Delta_{0,2}(0)$ = 0.49 meV [$\Delta_{0,2}(0)/k_{B}T_{c}$ = 2.57], are a bit larger and $x$ is smaller than the corresponding values obtained from the specific heat. This discrepancy might be attributed to various approximations of our modelling, such as the assumption of isotropy and the fact that we neglected interband scattering.  
Notwithstanding this slight discrepancy, our specific heat and penetration depth measurements coherently indicate the presence of two fully established, but significantly distinct superconducting gaps in La$_{7}$Ni$_{3}$.

\subsection{Zero-field $\mu$SR and broken TRS}
Zero-field muon spin relaxation (ZF-$\mu$SR) measurements were carried out to search for the presence of TRS breaking in La$_{7}$Ni$_{3}$. The ZF-$\mu$SR asymmetry spectra recorded below ($T$ = 0.2 K) and above ($T$ = 2.9 K) $T_{c}$ = 2.3 K are shown in \figref{Fig4:ZF}(a). The spectra do not exhibit any oscillatory components which rule out the presence of any long-range magnetically ordered state. The asymmetry spectra recorded below and above $T_{c}$ show an appreciable change in the relaxation rates, and the stronger relaxation rate below $T_{c}$ is indicative of the presence of internal magnetic fields in the superconducting state.

In the absence of atomic moments and at low temperature where muon diffusion is not appreciable, the relaxation rate is expected to arise from the local fields associated with the nuclear moments. In this case, the behaviour of asymmetry spectra is described by the Gaussian Kubo-Toyabe (KT) function \cite{Kubo} 
\begin{equation}
G_{\mathrm{KT}}(t) = \frac{1}{3}+\frac{2}{3}(1-\Delta^{2}t^{2})\mathrm{exp}\left(\frac{-\Delta^{2}t^{2}}{2}\right)
\label{eqn5:zf}
\end{equation}
where $\Delta$ represents the relaxation of muon spin due to the randomly oriented, static nuclear moments experienced at the muon site. The following function gives the best description of the ZF spectra for La$_{7}$Ni$_{3}$:
\begin{equation}
A(t) = A_{0}G_{\mathrm{KT}}(t)\mathrm{exp}(-\Lambda t)+A_{1}
\label{eqn6:tay}
\end{equation}
where $A_{0}$ is the initial sample-related asymmetry, $A_{1}$ is the background contribution to the asymmetry from the muons stopping in the sample holder whereas $\Lambda$ accounts for the electronic relaxation rates. 
The only parameter which exhibits a significant change as a function of temperature is the electronic relaxation rate, $\Lambda$, as shown in \figref{Fig4:ZF}(b), while $A_{0}$, $A_{1}$, and $\Delta$, see \figref{Fig4:ZF}(c), are found to be temperature independent. The magnetic field of 5mT was sufficient to decouple the muon spins from the relaxation channel, as shown in \figref{Fig4:ZF}(a). This points towards the presence of a static or quasi-static magnetic field and excludes the possibility of any extrinsic effects such as magnetic impurities. Such a notable increase in $\Lambda$ below the onset of superconductivity, suggests the appearance of a spontaneous magnetic field, providing convincing evidence of TRS breaking on entering the superconducting state of La$_{7}$Ni$_{3}$. 
Note that the signal corresponding to TRS breaking in La$_{7}$Ni$_{3}$ is similar to the one observed in other materials such as Sr$_{2}$RuO$_{4}$ \cite{SRO}, LaNiC$_{2}$ \cite{LNC}, SrPtAs \cite{SPA}, Lu$_{5}$Rh$_{6}$Sn$_{18}$ \cite{L5R6S18}, Y$_{5}$Rh$_{6}$Sn$_{18}$ \cite{Y5R6S18} and La$_{7}$Ir$_{3}$ \cite{L7I3_TRSB} which is considered to be a clear indication of broken TRS in these materials. %ascertain the TRS breaking in these materials. 
Our data also imply a dilute distribution of the sources of the TRS-breaking field. The increase in the relaxation rate below $T_{c}$ is $0.0081~\mu$s$^{-1}$ and the corresponding characteristic field strength is $\Lambda$/$\gamma_{\mu}$ = 0.1 G. Similar strength is reported for other compounds with superconducting states that break TRS \cite{L7R3_TRSB,LNC,SPA,L5R6S18,L7I3_TRSB}. 

Following $T_{c}$, there is first a decrease in $\Lambda$ upon lowering $T$ below the normal state value and the true increase in $\Lambda$ starts at $T_{\text{onset}} \approx$ 1.7 K. This leads to the small dip-like feature in \figref{Fig4:ZF}(b). We emphasize that there is no  indication for any second phase transition from other measurements such as TF $\mu$SR (penetration depth) and specific heat. However, this type of feature with $T_{\text{onset}}$ < $T_{c}$ has also been observed for other superconducting compounds such as in the odd-parity triplet state candidates PrPt$_{4}$Ge$_{12}$, Pr$_{1-x}$Ce$_x$Pt$_{4}$Ge$_{12}$, Pr$_{1-x}$La$_x$Pt$_{4}$Ge$_{12}$ \cite{PP4G12, PrCe, PrLa}, which are also believed to exhibit only a single superconducting transition.

\section{Discussion}
We next discuss the implications of these experimental findings for the microscopic form and origin of the superconducting phase in La$_{7}$Ni$_{3}$. We describe two possible microscopic scenarios separately.

\subsection{A single transition and $E_{2}(1,i)$ pairing}
\begin{figure}
\includegraphics[width=0.9\columnwidth,origin=b]{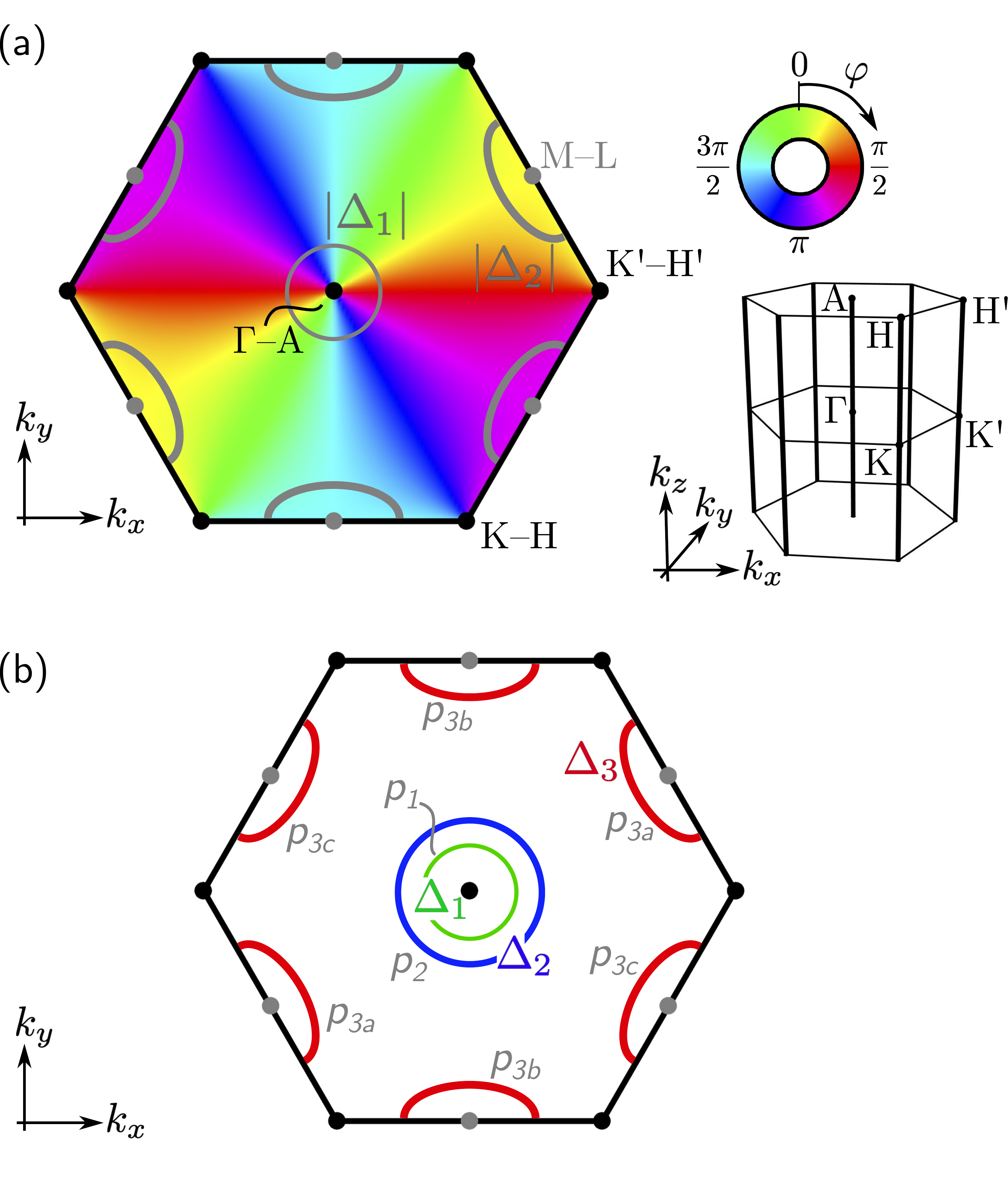}
\caption{\label{Fig5:Schematics} Schematic illustration of the two possible pairing scenarios, $E_2(1,i)$ and $s+i\,s$, consistent with experiment. (a) Phase of the order parameter (color) for the $E_2(1,i)$ state in a 2D cut (at fixed $k_z$) of the Brillouin zone for the lowest-order basis functions. As long as no Fermi surfaces (such as the examples shown in gray) go through the high-symmetry lines marked by black dots (thick black lines in the inset), the state is fully gapped. As indicated by the thickness of the gray lines, the gap in generals differs on symmetry-unrelated Fermi pockets, $|\Delta_1| \neq |\Delta_2|$, in agreement with experiment. For the $s+i\, s$ state, the TRS breaking results from non-trivial relative phases (indicated by colors) of the order parameter between pockets that are not related by symmetry, i.e., $\Delta_j/|\Delta_j|$ depends on $j$.}  
\end{figure}

Since there are currently no direct indications of two consecutive superconducting transitions, let us first assume that the TRS-breaking phase is reached by a single, continuous phase transition. Then the superconducting order parameter must transform under an irreducible representation (IR) of the point group, $C_{6v}$, of the crystal. Only multi-dimensional IRs, i.e., $E_1$ or $E_2$ for our case of $C_{6v}$, are consistent with the broken TRS.  
As already discussed in \refcite{L7R3_TRSB}, among the two associated superconductors, $E_{1,2}(1,i)$, that break TRS, $E_1(1,i)$ is a less natural candidate as it generically gives rise to line nodes, signs of which are not seen in our data. $E_{2}(1,i)$, however, can be gapless as long as no non-degenerate Fermi surfaces go through the high-symmetry lines indicated in the inset of \figref{Fig5:Schematics}(a) and will, otherwise, exhibit line nodes.  

To describe this pairing state microscopically, we first note that the broken inversion symmetry together with the presence of spin-orbit coupling removes the spin degeneracy of the bands. Therefore, spin is not a good quantum number at a generic crystal momentum $\vec{k}$ anymore and it is more natural \cite{DesignPrinciples} to study the projection, $\Delta(\vec{k})\in \mathbb{C}$, of the superconducting order parameter (multiplied by the unitary part of the time-reversal operator) onto the band closest to or at the Fermi level at $\vec{k}$. For the $E_{2}(1,i)$ pairing state, it has the form $\Delta(\vec{k}) = \chi_1(\vec{k}) + i\,\chi_2(\vec{k})$, where $\chi_1$ and $\chi_2$ are real-valued and Brillouin-zone-periodic basis functions transforming as $k_x^2 - k_y^2$ and $2k_xk_y$ under $C_{6v}$; $\chi_{1,2}(\vec{k})$ are further required to be even under $\vec{k}\rightarrow -\vec{k}$ as a consequence of Fermi statistics and TRS \cite{DesignPrinciples}. Here $k_{x,y}$ are momenta in the $ab$-plane of the system and perpendicular to two of the six mirror planes of $C_{6v}$. In \figref{Fig5:Schematics}(a), we show the resulting phase $\varphi_{\vec{k}}$ of the order parameter, $\Delta(\vec{k}) = e^{i \varphi_{\vec{k}}}|\Delta(\vec{k})|$, for the lowest-order (i.e., with fewest number of zeros) basis functions, $\chi_{1}(\vec{k})=\cos (k_x/2) \cos (\sqrt{3}k_y/2) - \cos k_x$, $\chi_{2}(\vec{k})=\sqrt{3}\sin (k_x/2) \sin (\sqrt{3}k_y/2)$, where we set $a=1$. We see that $e^{i \varphi_{\vec{k}}}$ is a smooth function, except for the $\Gamma$--A and the K$^{(\prime)}$--H$^{(\prime)}$ lines, where both $\chi_{1,2}$ and, thus, the gap vanish due to three-fold rotation symmetry, $C_3^z$, along $z$.

While we do not know the precise form of the Fermi surfaces of the system, we expect several pockets \cite{FirstPrinciples}. For illustration purposes, we show pockets (in gray) encircling both the $\Gamma$--A and the three M--L lines in \figref{Fig5:Schematics}(a). %, denoted by $p_4$ and $p_{1,2,3}$, respectively. 
The gap, given by $|\Delta(\vec{k})|$, is constrained by symmetry to be the same (and here approximated to be constant, $|\Delta(\vec{k})|=\Delta_2$) on the three disconnected sheets at the boundary of the Brillouin zone, enclosing the three symmetry related M--L lines. It is generically not equal to that ($\Delta_1$) in the Brillouin-zone center; $\Delta_1 \neq \Delta_2$ will give rise to the two-gap behavior seen in the specific heat and penetration depth. Note that, in this scenario, the $\vec{k}$-dependent phase that breaks TRS occurs between parts of the Fermi surface that are related by symmetry, such as the three pockets at the Brillouin-zone boundary which are related by $C_6^z$ and exhibit a relative phase of (approximately) $e^{2\pi i/3}$.

Such a pairing state is expected to be realized, for instance, when the dominant interaction in the Cooper channel is a repulsion between the three pockets at the Brillouin-zone boundary in \figref{Fig5:Schematics}(a); we refer to the toy model in \appref{ToyModel} for an illustration. In fact, it was very generally shown \cite{SheurerGenRelation} that conventional electron-phonon pairing alone cannot give rise to a TRS-breaking superconducting state. Instead, an unconventional mechanism is required, associated with the fluctuation of a time-reversal-odd collective electronic mode, such as spin-fluctuations.

\subsection{Two consecutive transitions and $s+i\,s$ pairing}
% The possibility of two transitions:
A totally different route to TRS-breaking, full-gap superconductivity proceeds by two consecutive transitions, which could be too close to be resolved experimentally. Such a scenario arises when the dominant repulsive Cooper-channel interactions in the system scatter Cooper pairs of symmetry-\textit{un}related parts of the Fermi surface---for instance between the three symmetry-unrelated sets of sheets $p_1$, $p_2$, and $p_{3}=p_{3a} \cup p_{3b} \cup p_{3c}$ in \figref{Fig5:Schematics}(b). Let us assume that $\Delta_{\vec{k}}$ only varies weakly on each of the pockets and approximate $\Delta_{\vec{k}} = \Delta_j$ if $\vec{k}\in p_j$. The system is  ``frustrated'' in the sense that these interactions favor $\Delta_1=-\Delta_2$, $\Delta_1=-\Delta_3$, but also $\Delta_2=-\Delta_3$ simultaneously. If these dominant interactions are of the same order, the best compromise consists of non-trivial relative complex phases between $\Delta_{1,2,3}$, breaking TRS and shown as differently colored Fermi pockets in \figref{Fig5:Schematics}(b). The resulting fully-gapped $s+ i\, s$ state, with complex phases between symmetry-\textit{un}related parts of the Fermi surface, cannot be reached by a single second-order phase transition as it transforms trivially under $C_{6v}$ (IR $A_1$) and yet breaks TRS. Instead, there have to be two consecutive transitions and $\Delta(\vec{k}) = \eta_{s_1}(T) \chi_{s_1}(\vec{k}) + \eta_{s_2}(T) \chi_{s_2}(\vec{k})$, with $\chi_{s_j}(g\vec{k})=\chi_{s_j}(\vec{k})\in\mathbb{R}$ for all $g\in C_{6v}$ but changing sign between different $p_j$; at $T_{c1}$, first $\eta_{s_1} \neq 0$ while $\eta_{s_2}$ only becomes finite below $T_{c2}<T_{c1}$, with a non-trivial complex phase relative to $\eta_{s_1}$. The two temperatures can get very close if the different dominant interaction scales (times the respective density of states) are close to each other. We further illustrate this in \appref{ToyModel} and also point out that it is conceptually related to a proposed mechanism of TRS breaking in the (centrosymmetric) iron-based superconductors \cite{Andrey}.

\subsection{Relevance of disorder}
% Finally, a few comments on disorder:
Finally, our measurements indicate (see \appref{ExtractParameter}) that the electronic mean-free path $l$ of our sample of La$_{7}$Ni$_{3}$ is of the same order as the superconducting coherence length $\xi$. One would, thus, naively expect that any superconducting state with a significantly momentum-dependent order parameter, such as the $\vec{k}$-dependent phases of $\Delta(\vec{k})$ above, should be strongly suppressed \cite{BW,GolubovMazin}. However, more recent theoretical works \cite{DisorderSOCFu,OurDisorderSOC,BrydonScattering,PdTeScattering,Jonathan} have shown that the pair-breaking effect of impurities can be significantly reduced in the presence of spin-orbit coupling; this is also confirmed by experiments \cite{Ando2012,Ando2014,Welp} indicating that the critical temperature of pairing states transforming under a non-trivial representation of the point group might only be weakly suppressed in the presence of spin-orbit coupling---even when $l$ becomes smaller than $\xi$ (or even of order of the Fermi wavelength). The basic reason for this protection mechanism is that matrix elements of an impurity potential between certain pairs $(\vec{k}_1,\vec{k}_2)$ of momenta can be suppressed due to spin-orbit mixing and that only certain scattering processes are pair breaking. As follows immediately from the generalized Anderson theorem of \cite{Scheurer2016,Hoyer2015,PdTeScattering} (see \appref{ToyModel} for more details), the $E_1(1,i)$ state in \figref{Fig5:Schematics}(a), for instance, is not affected by scattering within each of the three pockets at the zone boundaries. %, i.e., both $\vec{k}_1$ and $\vec{k}_2$ are on the same $\{p_{2a},p_{2b},p_{2c}\}$. 
Similarly, the $s+i\, s$ state is protected against scattering events where both $\vec{k}_1$ and $\vec{k}_2$ belong to one of the three sets of sheets $p_1$, $p_2$, and $p_{3}$. We, thus, believe that systematic combined experimental and theoretical studies of the impact of disorder can provide further information about the form of the superconducting pairing state.

\section{Summary and conclusion}
In summary, we have studied the superconducting state of La$_{7}$Ni$_{3}$, the most recent member of the Th$_{7}$Fe$_{3}$ family of superconductors, using a combination of specific heat, TF-$\mu$SR, and ZF-$\mu$SR measurements. Both specific heat and TF measurements reveal the presence of two significantly distinct superconducting gaps, which are both found to be nodeless and isotropic. In ZF measurements, we have detected a static internal magnetic field that starts to develop below the superconducting transition temperature and grows as we further decrease temperature; this provides substantial evidence for TRS breaking and, hence, reveals the unconventional nature of the superconducting state of La$_{7}$Ni$_{3}$. The change in the relaxation rate for La$_{7}$Ni$_{3}$ is almost equivalent to other compounds such as La$_{7}$Ir$_{3}$ and La$_{7}$Rh$_{3}$. This points towards the relevance of La in inducing the unconventional superconducting state in this family of materials. % and warrants study of elemental La metal. 
We have discussed possible microscopic scenarios for the order parameter and pairing interactions of the superconducting phase. In order to pinpoint, which of those possibilities is realized, further experimental work on single crystals of La$_{7}$Ni$_{3}$ and band structure calculations are required.

\vspace{2em}

\begin{acknowledgments}
R.~P.~S.\ acknowledges the Science and Engineering Research Board, Government of India for the Core Research Grant CRG/2019/001028. Arushi acknowledges the funding agency, University Grant Commission (UGC) of Government of India for providing SRF fellowship. We thank ISIS, STFC, UK for the Newton funding and beamtime to conduct the $\mu$SR experiments. M.~S.~S. acknowledges discussions with P.~Orth.
\end{acknowledgments}

%\pagebreak
%======================================================================================================
\appendix

\section{Parameters of La$_{7}$Ni$_{3}$ from data}\label{ExtractParameter}
\subsection{Critical fields}
\begin{figure} %{r}{0.5\textwidth}
\includegraphics[width=1.0\columnwidth]{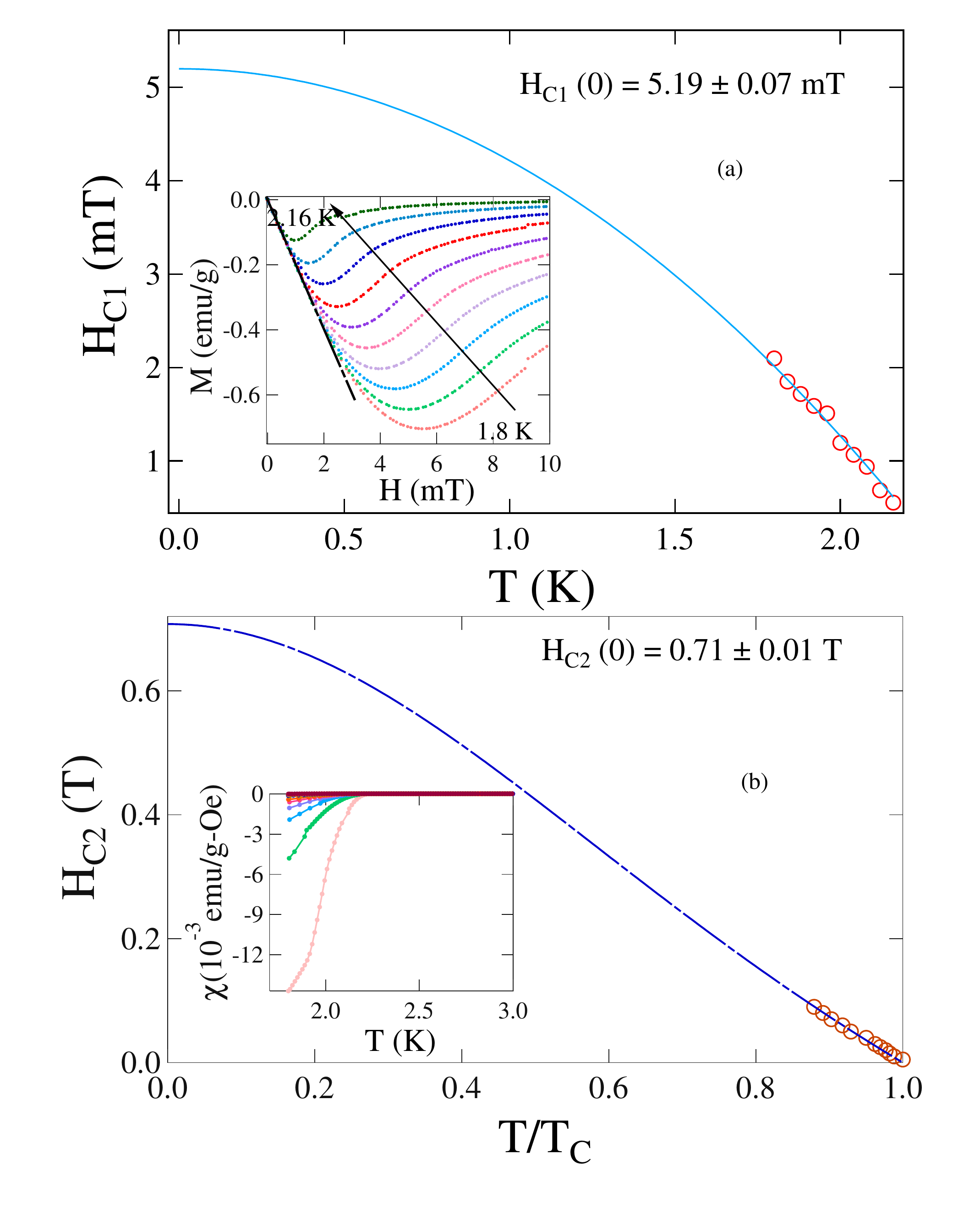}
\caption{ \label{Fig1:CF} (a) Lower critical field, $H_{c1}$, as a function of temperature. Inset shows the low-field magnetization curves obtained at different fixed temperatures. (b) Upper critical field, $H_{c2}$, of La$_{7}$Ni$_{3}$ versus temperature, where the blue line represents a fit with Eq.~\eqref{eqn2:Hc2}. Inset shows the susceptibility as a function of temperature at different fields.}
\end{figure}

To determine the lower critical field, $H_{c1}(T=0)$, field-dependent magnetization, $M(H)$, measurements have been performed with $T$ ranging from 1.8~K to 2.16~K, as shown in the inset of \figref{Fig1:CF}(a). $H_{c1}$ is defined as the value of $H$ where  $M(H)$ starts to deviate from linearity. The main panel of \figref{Fig1:CF}(a) shows the estimated value of $H_{c1}$ as a function of $T$. The solid blue line is a fit to the data using the following equation:
\\
\begin{equation}
H_{c1}(T)=H_{c1}(0)\left[1-\left(\frac{T}{T_{c}}\right)^{2}\right].
\label{eqn1:Hc1}
\end{equation} 
\\
It provides $H_{c1}(0) = 5.19 \pm 0.07\, \textrm{mT}$. 

The upper critical field has been extracted from the $\chi(T)$ curves in the inset of \figref{Fig1:CF}(b), measured at different magnetic fields starting from 5 mT to 90 mT by considering the onset of the diamagnetic signal. %as the transition temperature, T$_{C}$. 
Figure~\ref{Fig1:CF}(b) shows a linear behaviour in $H_{c2}(T)$ when plotted with respect to temperature and can be well described using the Ginzburg-Landau equation
\\
\begin{equation}
H_{c2}(T) = H_{c2}(0)\frac{1-t^{2}}{1+t^2}, 
\label{eqn2:Hc2}
\end{equation} 
\\
where $t = T/T_{c}$. The data fits well and provides $H_{c2}(0) = 0.71 \pm 0.01\,\textrm{T}$. The Pauli paramagnetic limit is given by $ H_{c2}^{P}(0) = C\,T_{c}$, where $C = 1.86\, T/K$. It is evaluated to be $4.28 \,\textrm{T}$ which is higher than the value of the upper critical field and indicates the dominance of spin singlet component in the superconducting state. 

The value of $H_{c2}(0)$ is also used to determine the Ginzburg-Landau coherence length $\xi_{\text{GL}}(0)$ by using the expression \cite{Coh_Leng&pene}: $H_{c2}(0) = \frac{\Phi_{0}}{2\pi\xi_{\text{GL}}^{2}}$ where $ \Phi_{0}$  (= 2.07 $\times$ 10$^{-15}$ Tm$^{2}$) is the superconducting magnetic-flux quantum. For $H_{c2}(0) = 0.71 \,T$, $\xi_{\text{GL}}$(0) is evaluated to be 215 $\text{\AA}$. In order to calculate the Ginzburg-Landau penetration depth, $\lambda_{\text{GL}}$(0), the values of $\xi_{\text{GL}}$(0) and $H_{c1}(0)$ can be employed in the relation \cite{Coh_Leng&pene}:
\begin{equation}
H_{c1}(0) = \frac{\Phi_{0}}{4\pi\lambda_{\text{GL}}^2(0)}\left(\mathrm{ln}\frac{\lambda_{\text{GL}}(0)}{\xi_{\text{GL}}(0)}+0.12\right)   
\label{eqn4:PD}
\end{equation} 
With  $H_{c1}(0) = 5.19 \,\textrm{mT}$ and $\xi_{\text{GL}}$(0) = 215 $\text{\AA}$, we find $\lambda_{\text{GL}}(0) = 2940 \,\text{\AA}$. Consequently, the Ginzburg-Landau parameter, $\kappa_{\text{GL}} = \frac{\lambda_{\text{GL}}(0)}{\xi_{\text{GL}}(0)} = 26$ suggests that La$_{7}$Ni$_{3}$ is a strong type II superconductor. All the calculated parameters are summarized in Table~\ref{TableWithParameters} together with its isostructural compounds La$_{7}$Rh$_{3}$ and La$_{7}$Ir$_{3}$.

\subsection{Electrical Resistivity}
\begin{figure} %{r}{0.5\textwidth}
\includegraphics[width=1.0\columnwidth]{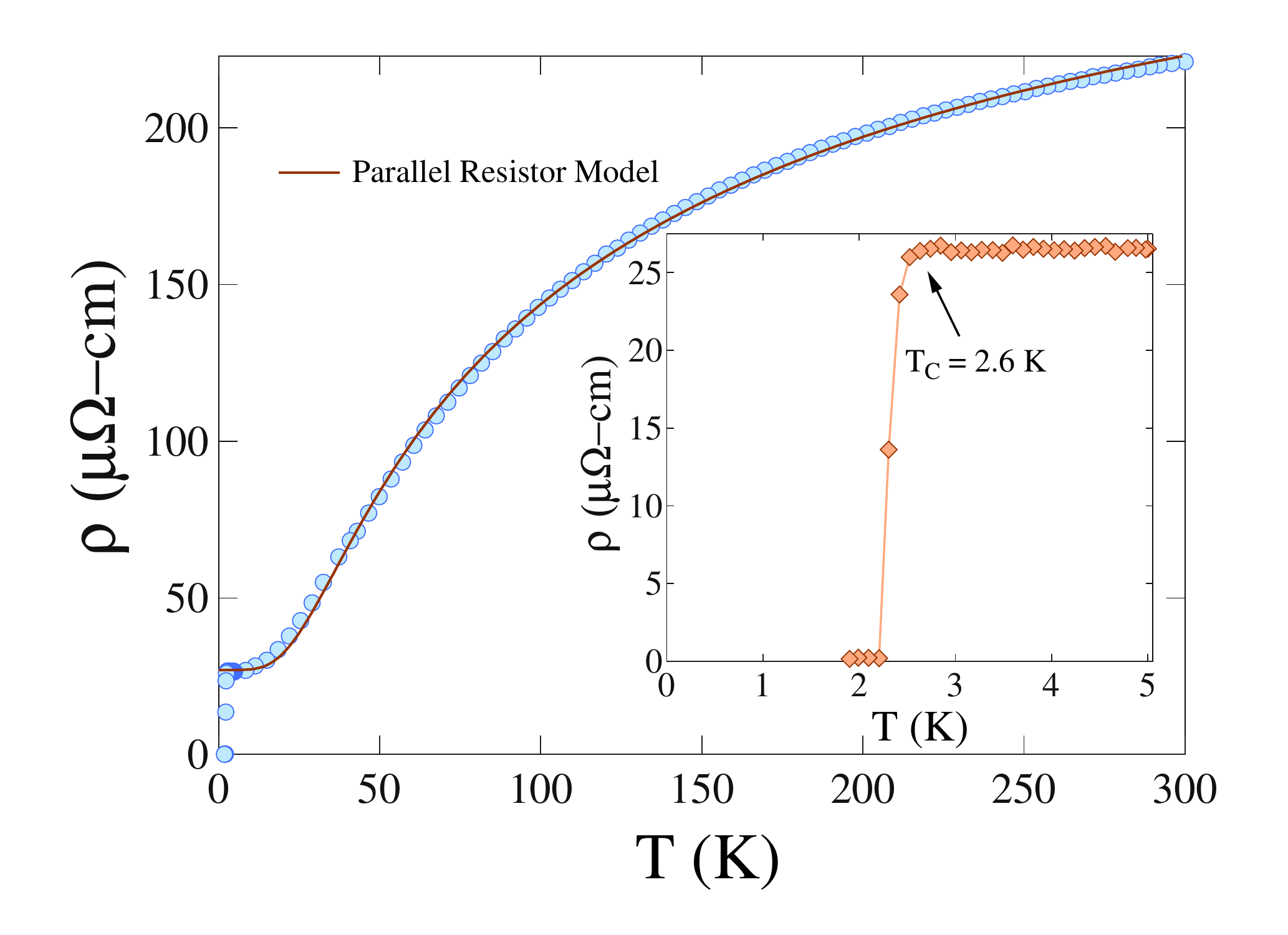}
\caption{ \label{Fig2:res} Resistivity as a function of temperature in zero applied field over the range 1.9 K $\leq$ T $\leq$ 300 K. The line shows a fitting with parallel resistor model. Inset: $\rho(T)$ at low-temperature exhibiting a drop in resistivity at $T_{c,\text{onset}} = 2.6 K$.}
\end{figure}

Electrical resistivity measurements in the temperature range 1.9 K to 300 K have been performed in zero applied field and are shown in \figref{Fig2:res}. Inset shows the zero-resistivity drop with onset temperature $T_{c,\text{onset}} = 2.6\,\text{K}$. Above $T \approx 50\,\textrm{K}$, the resistivity data exhibits a saturation-like behavior which can occur when the mean free path is of the order of the inter-atomic spacing \cite{InterSpac}. This type of behavior in $\rho(T)$ was described by Wiesmann \textit{et al.} \cite{Parallel} by the following expression
\begin{equation}
 \rho(T) = \left[\frac{1}{\rho_{s}} + \frac{1}{\rho_{i}(T)} \right]^{-1}
\label{para1}
\end{equation}
where $\rho_{s}$ is the temperature-independent saturation resistivity attained at higher temperatures and $\rho_{i}(T)$ is given by \cite{BG}
\begin{equation}
 \rho_{i}(T) = \rho_{i,0} + C\left(\frac{T}{\Theta_{D}}\right)^{5}\int_{0}^{\Theta_{D}/T}\frac{x^{5}}{(e^{x}-1)(1-e^{-x})}dx
\label{para2}
\end{equation}
where $\rho_{i,0}$ represents the temperature-independent residual resistivity due to scattering from defects in the crystal structure; the second term in \equref{para2} is a temperature-dependent contribution which can be described by the generalized BG model in which $C$ is a material-dependent factor and $\Theta_{D}$ is the Debye temperature obtained from resistivity measurements. The red dashed line in main panel of \figref{Fig2:res} shows the best fitting to the data and yields $\rho_{0}$ = 29.6 $\pm$ 0.3 $\mu\Omega $ cm, $C$ = 1674 $\pm$ 39 $\mu\Omega $ cm, $\rho_{0,s}$ = 303 $\pm$ 1 $\mu\Omega $ cm, and $\Theta_{D}$ = 152 $\pm$ 2 K. 

\begin{table}[tb]
\caption{Parameters obtained from various measurements.}
%\begin{center}
%\begin{tabular*}{1.049\columnwidth}{l@{\extracolsep{\fill}}lcccr}\hline
%\end{tabular*}
%\end{center}

%\begingroup
%\setlength{\tabcolsep}{8.2pt}
\vspace{0.4em}
\label{TableWithParameters}
\begin{ruledtabular}
\begin{tabular}[b]{lcccc}
Parameters&  Units& La$_{7}$Ni$_{3}$& La$_{7}$Rh$_{3}$ \cite{L7R3_TRSB}& La$_{7}$Ir$_{3}$ \cite{L7I3}\\
\hline
$T_{c}$& K& 2.3& 2.65& 2.25\\ 
$H_{c1}$(0)& mT& 5.19& 2.51& 3\\
$H_{c2}$(0)& T& 0.71& 1.02& 0.97\\
$\xi_{\text{GL}}$(0)& \text{\AA}& 215& 179& 174\\
$\lambda_{\text{GL}}$(0)& \text{\AA}& 2940& 4620& 4720\\
$\kappa_{\text{GL}}$& & 26& 26& 21
\\[0.1ex]
\end{tabular}
\end{ruledtabular}
\par\medskip\footnotesize
%\endgroup
\end{table}

\subsection{Electronic Properties}
%In order to determine the clean or dirty limit superconductivity, a set of equations has been used which calculate the BCS coherence length and mean free path. 
The Sommerfeld coefficient, $\gamma_n$, and  quasiparticle number density, $n$, %per unit volume 
are related by
\begin{equation}
\gamma_{n} = \left(\frac{\pi}{3}\right)^{2/3}\frac{k_{B}^{2}m^{*}V_{\mathrm{f.u.}}n^{1/3}}{\hbar^{2}N_{A}}
\label{eqn17:gf}
\end{equation}
where $k_{B}$ is the Boltzmann constant, V$_{\mathrm{f.u.}}$ is the volume of a formula unit, $N_{A}$ is the Avogadro number and $m^{*}$ is the effective mass of quasiparticles. The residual resistivity $\rho_0$ is related to Fermi velocity $v_{\mathrm{F}}$ and electronic mean free path $\textit{l}$ by the expression
 \begin{equation}
\textit{l} = \frac{3\pi^{2}{\hbar}^{3}}{e^{2}\rho_{0}m^{*2}v_{\mathrm{F}}^{2}}
\label{eqn18:le}
\end{equation}
whereas the Fermi velocity $v_{\mathrm{F}}$ can be expressed in terms of quasiparticle carrier density and effective mass by
\begin{equation}
n = \frac{1}{3\pi^{2}}\left(\frac{m^{*}v_{\mathrm{F}}}{\hbar}\right)^{3} .
\label{eqn19:n}
\end{equation}
The expression for the penetration depth $\lambda_{\text{GL}}$(0) in the dirty limit is given by \cite{MTinkham}
\begin{equation}
\lambda_{\text{GL}}(0) = \left(\frac{m^{*}}{\mu_{0}n e^{2}}\right)^{1/2}\left(1+\frac{\xi_{0}}{\textit{l}}\right)^{1/2}
\label{eqn20:f}
\end{equation}
where $\xi_{0}$ is the BCS coherence length and the first term in the bracket represents the London penetration depth $\lambda_{L}$. In the dirty limit at T = 0 K, Ginzburg-Landau coherence length $\xi_{\text{GL}}$(0) and BCS coherence length $\xi_{0}$ are related by the expression
\begin{equation}
\frac{\xi_{\text{GL}}(0)}{\xi_{0}} = \frac{\pi}{2\sqrt{3}}\left(1+\frac{\xi_{0}}{\textit{l}}\right)^{-1/2}.
\label{eqn21:xil}
\end{equation}
By providing the values of $\gamma_{n}$ = 44.44 mJmol$^{-1}$K$^{-2}$, $\xi_{\text{GL}}$ (0) = 215 $\text{\AA}$, and $\rho_{0}$ = 29.6 $\mu\Omega$-cm, the above equations were solved simultaneously which provides m$^{*}$ = 6.3 m$_{e}$, $n$ = 4.3$\times$10$^{27}$ m$^{-3}$, $l_{e}$ = 182 $\text{\AA}$, $\xi_{0}$ =  202 $\text{\AA}$. The ratio of $\xi_{0}$/$l_{e}$ = 1.1 indicates that the superconductor is characterized by significant disorder (``dirty limit''), which is consistent with published data \cite{L7N3_macro}.

\section{Toy models for TRS breaking}\label{ToyModel}
In this appendix, we will discuss symmetry-based, phenomenological models for each of the two possible TRS-breaking pairing scenarios---the $E_2(1,i)$ and $s+i\,s$ states---discussed in the main text. We will also briefly study their disorder sensitivity, using the results of \cite{Scheurer2016,PdTeScattering}.

\subsection{$E_2(1,i)$ pairing model}
To provide an example of how an $E_2(1,i)$ state can arise, let us assume that the energetics of superconductivity is mainly determined by the interactions between the three distinct pockets encircling the M--L lines in \figref{Fig5:Schematics}(a) and that the superconducting order parameter can be taken to be constant on each of these three separate sheets, $j=1,2,3$, with value $\Delta(\vec{k})=\Delta_j$. Neglecting the subleading energetic contributions from other Fermi surfaces for simplicity, the superconducting free-energy can be written as
\begin{equation}
    \Delta\mathcal{F}(T) = \sum_{j,j'=1}^3\Delta_j^* a_{jj'}(T) \Delta_{j'} + \dots , \label{GeneralFreeEnergyExpansion}
\end{equation}
where the ellipsis stands for terms higher order in the superconducting order parameter.
It is straightforward to derive the representations of the symmetries on the superconducting order parameter: under time reversal, $\Delta_j \rightarrow \Delta_j^*$; under $C_6^z$, $\Delta_j \rightarrow \Delta_{j+1}$ (with $\Delta_4 \equiv \Delta_1$), and under $\sigma_v$, $\vec{\Delta}:=(\Delta_1,\Delta_2,\Delta_3) \rightarrow (\Delta_3,\Delta_2,\Delta_1)$. Combined with $\Delta\mathcal{F} \in \mathbb{R}$, we thus end up with
\begin{equation}
    a_{jj'}(T) = \alpha(T) \delta_{j,j'} + \beta(T) (1- \delta_{j,j'}), \quad \alpha,\beta\in\mathbb{R},
\end{equation}
such that the dominant superconducting instability (eigenvector of $a(T)$ with eigenvalue that first becomes negative upon lowering $T$ to $T_c$) just depends on the sign of $\beta(T_c)$: if it is negative, which corresponds to \textit{attractive inter-pocket interactions}, we get the $s$-wave state (transforming under IR $A_1$ of $C_{6v}$)
\begin{equation}
    \vec{\Delta} = \Delta_0(T) (1,1,1), \qquad \Delta_0(T)\in\mathbb{C},
\end{equation}
which preserves TRS. For \textit{repulsive interpocket interactions}, $\beta(T_c)>0$, the leading instability is doubly degenerate,
\begin{align*}
    \vec{\Delta} = \sum_{\mu=1,2} \eta_\mu \vec{\chi}_{\mu},& \quad \eta_\mu\in\mathbb{C}, \\
    \vec{\chi}_1 = \left(\frac{1}{\sqrt{6}},-\sqrt{\frac{2}{3}},\frac{1}{\sqrt{6}}\right), \quad &\vec{\chi}_2 = \left(\frac{1}{\sqrt{2}},0,-\frac{1}{\sqrt{2}}\right),
\end{align*}
with the two components $\vec{\chi}_{1,2}$ transforming as $k_x^2 - k_y^2$ and $2k_xk_y$ under $C_{6v}$ and, hence, under the IR $E_2$.
While the two complex numbers $\eta_{1,2}$ are determined by the higher-order terms omitted in \equref{GeneralFreeEnergyExpansion}, it has been shown \cite{DesignPrinciples} that, irrespective of microscopic details, only the solution with a non-trivial relative complex phase, $\eta_{2}=\pm i\eta_1$, is possible energetically for spin-orbit coupled noncentrosymmetric systems. Therefore, we end up with
\begin{equation}
    \vec{\Delta} = \, \Delta_0(T)\, (1,\omega,\omega^2), \qquad \omega = e^{\frac{2\pi i}{3}}, \label{FinalFormOfSCOrderparameter}
\end{equation}
in accordance with the form of the complex phase shown in \figref{Fig5:Schematics}(a). 

To demonstrate the statements of the main text about which scattering processes are pair breaking, we employ the generalized form of the Anderson theorem of \refcite{Scheurer2016} (see Sec.~7.1 therein): for our case of singly-degenerate Fermi surfaces, a superconductor is protected against non-magnetic, $t_W=+1$, (magnetic, $t_W=-1$) impurity scattering with matrix elements $W_{\vec{k},\vec{k}'}$ if the commutator (anti-commutator)
\begin{equation}
    C_{\vec{k},\vec{k}'} = \Delta(\vec{k})W_{\vec{k},\vec{k}'} -t_W W_{\vec{k},\vec{k}'} \Delta(\vec{k}') \label{FormOfTheCommutator}
\end{equation}
vanishes. In fact, it was shown \cite{PdTeScattering} that the same commutator also quantitatively determines how fragile a superconductor is, since the reduction $\delta T_c = T_c - T_{c,0}$ of the transition temperature $T_c$ relative to its value in the clean limit $T_{c,0}$ can be written as
\begin{equation}
    \delta T_c \sim -\frac{\pi}{4} \tau^{-1} \zeta
\end{equation}
for small total scattering rates $\tau^{-1} \rightarrow 0$. The sensitivity parameter reads as \cite{PdTeScattering}
\begin{equation}
    \zeta = \frac{||C||^2_{\text{F}}}{4 ||W||^2_{\text{F}} \braket{|\Delta|^2}_{\text{FS}}}, \label{FormOfZeta}
\end{equation}
where $||C||^2_{\text{F}} = \sum^{\text{FS}}_{\vec{k},\vec{k}'} |C_{\vec{k},{\vec{k}'}}|^2$ is Frobenius norm on the Fermi surface (FS) and $\braket{\cdot}_{\text{FS}}$ is the Fermi surface average (normalized such that $\braket{1}_{\text{FS}}=1$). 

This expression is readily applied to any superconducting system and impurity potential \cite{PdTeScattering,OurTBGApplication}. For our model here, with order parameter given in \equref{FinalFormOfSCOrderparameter}, the scattering matrix $W$ effectively becomes a Hermitian $3\times 3$ matrix, $W_{jj'}=W^*_{jj'}$, $j,j'=1,2,3$. After straightforward matrix algebra, \equref{FormOfZeta} yields 
\begin{equation}
    \zeta_{E_2(1,i)} = \frac{3\tau^{-1}_{\text{inter}}}{4(\tau^{-1}_{\text{intra}} + \tau^{-1}_{\text{inter}})}, \label{ZetaE2}
\end{equation}
with inter- and intra-sheet scattering rates given by $\tau^{-1}_{\text{inter}} = 2(|W_{12}|^2 + |W_{13}|^2 + |W_{23}|^2)$ and $\tau^{-1}_{\text{intra}} = \sum_{j=1}^3 |W_{jj}|$, respectively. In particular, it follows that only inter-sheet scattering is pair-breaking, as stated in the main text. This can also be directly seen from the commutator (\ref{FormOfTheCommutator}) of the generalized Anderson theorem, which vanishes for momentum pairs $(\vec{k},\vec{k}')$ with $\Delta_{\vec{k}}=\Delta_{\vec{k}'}$ in the current case of non-magnetic scattering ($t_W=1$). Further note we get $\zeta=1/2$ from \equref{ZetaE2} for momentum-independent scattering matrix elements, $|W_{ij}|=\text{const.}$, as expected for a pairing state transforming under a non-trivial IR.

\subsection{$s+i\,s$ pairing model}
To illustrate how an $s+i\,s$ state can arise, let us consider three different sets of symmetry-unrelated Fermi pockets, such as $p_1$, $p_2$, and $p_{3}=p_{3a} \cup p_{3b} \cup p_{3c}$ in \figref{Fig5:Schematics}(b). Since we are interested in $s$-wave pairing, we will assume that $\Delta(\vec{k})$ is invariant under all $g\in C_{6v}$ and, for simplicity, take it to be constant on each of these three sets of sheets, $\Delta(\vec{k}) = \Delta_j$ if $\vec{k} \in p_j$. The free-energy can again be written in the form (\ref{GeneralFreeEnergyExpansion}) to quadratic order in $\Delta$; however, since the pockets are not related by symmetry operators, there are fewer constraints on $a(T)$: it only needs to be real and symmetric, $a^T=a\in\mathbb{R}^{3\times 3}$, due to $\Delta\mathcal{F} \in \mathbb{R}$ and TRS. Further assuming that the form of the pairing state is predominantly determined by the repulsion between the different pockets, we can use
\begin{equation*}
    a(T) \approx \alpha_0(T) \mathbbm{1}_{3\times 3} + \beta(T) \begin{pmatrix}0 & 1 & 1 + \delta_1 \\ 1 & 0 & 1 +\delta_2 \\ 1 + \delta_1 & 1 + \delta_2 & 0 \end{pmatrix}, 
\end{equation*}
with $\beta(T) < 0$, $\delta_j > -1$. As such, the form of the leading instability only depends on the two real parameters $\delta_{1,2}$, which we rewrite according to $(\delta_1,\delta_2) = \delta (\cos\theta,\sin\theta)$. For $|\delta| \ll 1$, two superconducting states are asymptotically degenerate and 
\begin{equation}
    \vec{\Delta} = (\Delta_1,\Delta_2,\Delta_3)^T = \sum_{\mu=\pm}\eta_\mu(T) \vec{\chi}_\mu, \quad \eta_\mu\in\mathbb{C},
\end{equation}
where 
\begin{equation}
    \vec{\chi}_{\pm} \sim \mathcal{N}_{\pm}(\theta) \begin{pmatrix} \cos\theta - \sin\theta \pm\sqrt{1 -\cos\theta \sin\theta} \\ \sin\theta \\ \mp \sqrt{1 -\cos\theta \sin\theta} - \cos\theta   \end{pmatrix}, \label{FormOfChiPM}
\end{equation}
with normalization factor $\mathcal{N}_{\pm}(\theta)$. As follows from the respective eigenvalues, $\eta_+$ ($\eta_-$) first becomes non-zero at $T_{c1}$ upon cooling down while $\eta_-=0$ ($\eta_+=0$), if $\delta> 0$ ($\delta<0$). This first superconducting phase respects TRS since $\vec{\Delta}^* \propto \vec{\Delta}$. However, due to the near degeneracy of the states, a second transition can take place at which $\eta_-\neq 0$ ($\eta_+\neq 0$), with transition temperature $T_{c2}$; note that $T_{c1}-T_{c2}$ becomes vanishingly small as $\delta \rightarrow 0$.

To obtain the relative weight and phase of $\eta_{+}$ and $\eta_{-}$, we need to go beyond quadratic order in $\Delta\mathcal{F}$. Among the quartic contributions to the free energy, only 
\begin{equation}
    \gamma(T) \text{Re}\left[ (\eta_+^*)^2 \eta_-^2 \right]
\end{equation}
is sensitive to the relative phase. It straightforwardly follows from the structure of the underlying one-loop diagram that $\gamma(T) > 0$, see, e.g., \refcite{Andrey} for an explicit computation. Therefore, we obtain that the relative complex phase of $\eta_+$ and $\eta_-$ is $\pi/2$ below $T_{c2}$. Since $\vec{\chi}_{\pm}$ in \equref{FormOfChiPM} are linearly independent, the superconducting state breaks TRS as stated in the main text. 

As follows from this discussion, the $s+i\,s$ state is possible as long as neither $\vec{\Delta} = \vec{\chi}_+$ nor $\vec{\Delta} = \vec{\chi}_-$ are suppressed by disorder. While it is straightforward to evaluate \equref{FormOfZeta} for the general form in \equref{FormOfChiPM}, let us for notational simplicity focus on $\delta_1=\delta_2\equiv \delta$ where $\vec{\chi}_+ \sim (1+\delta/3,1+\delta/3,-2)^T$, to linear order in $\delta$, and $\vec{\chi}_- \propto (-1,1,0)^T$. For $\vec{\chi}_-$, we then get
\begin{equation}
    \zeta_- = \frac{3(\tau^{-1}_{\text{inter}}/2 + 3 |W_{12}|^2)}{4(\tau^{-1}_{\text{intra}} + \tau^{-1}_{\text{inter}})},
\end{equation}
and $\vec{\chi}_+$ leads to
\begin{equation}
    \zeta_+ = \frac{(9 + \delta)^2}{27 + \delta(6+\delta)} \frac{3(|W_{13}|^2+|W_{23}|^2)}{4(\tau^{-1}_{\text{intra}} + \tau^{-1}_{\text{inter}})} .
\end{equation}
As expected, we see that, while $\vec{\chi}_-$ is particularly sensitive to scattering between the pockets $p_1$ and $p_2$, only scattering events between patches $p_1$ and $p_3$, and between $p_2$ and $p_3$ are pair breaking for $\vec{\chi}_+$, which can also be immediately inferred from the commutator (\ref{FormOfTheCommutator}) directly. It is thus manifest that intra-sheet scattering is not expected to suppress the $s+i\,s$ state.

\end{document}